\documentclass[sigconf]{acmart}

\usepackage{soul}
\usepackage{hyperref}
\usepackage{multirow}
\usepackage{listings}
\usepackage{fancybox}
\usepackage{enumitem}
\usepackage{graphicx}
\usepackage{color}
\usepackage{fancyvrb}
\usepackage{array}
\usepackage{amsmath}
\usepackage{xspace}
\usepackage{url}
\usepackage{tikz}
\usepackage{caption}
\usepackage{pgfplots}
\usepackage{subcaption}
\pgfplotsset{compat=1.16}
\usepackage{pgf-pie}
\usepackage{float}
\usetikzlibrary{pgfplots.statistics,calc}
\usepackage{algorithm}
\usepackage{algorithmic}
\usepackage{adjustbox}
\usepackage{makecell}

\makeatletter
\newcommand{\mybox}[1]{%
	\setbox0=\hbox{#1}%
	\setlength{\@tempdima}{\dimexpr\wd0+13pt}%
	\begin{tcolorbox}[boxrule=0.5pt, colback=white, arc=4pt,
		left=6pt,right=6pt,top=6pt,bottom=6pt,boxsep=0pt]
		#1
	\end{tcolorbox}
}

\definecolor{songcolor}{RGB}{191,191,191}

\newcommand{\tool}{\texttt{PAGENT}}

\AtBeginDocument{%
  }

\setcopyright{acmcopyright}
\copyrightyear{2024}
\acmYear{2024}
\acmDOI{XXXXXXX.XXXXXXX}
\acmPrice{15.00}
\acmISBN{978-1-4503-XXXX-X/18/06}
\usepackage{listings}
\usepackage{xcolor}
\usepackage{balance}
\usepackage{tcolorbox}
\newcounter{o}
\begin{document}

\definecolor{codegreen}{rgb}{0,0.6,0}
\definecolor{codegray}{rgb}{0.5,0.5,0.5}
\definecolor{codepurple}{rgb}{0.58,0,0.82}
\definecolor{backcolour}{rgb}{0.95,0.95,0.92}

\lstdefinestyle{mystyle}{
    backgroundcolor=\color{backcolour},   
    commentstyle=\color{codegreen},
    keywordstyle=\color{magenta},
    numberstyle=\tiny\color{codegray},
    stringstyle=\color{codepurple},
    basicstyle=\ttfamily\footnotesize,
    breakatwhitespace=false,         
    breaklines=true,                 
    captionpos=b,                    
    keepspaces=true,                 
    numbers=left,                    
    numbersep=5pt,                  
    showspaces=false,                
    showstringspaces=false,
    showtabs=false,                  
    tabsize=2
}

\lstset{style=mystyle}

\title[]{PAGENT: Learning to Patch Software Engineering Agents}

\begin{abstract}
LLM Agents produce patches automatically to resolve an issue. 
However, they can generate inaccurate patches. Little is known about the root causes behind those failed patches or how those could be fixed. In this paper, we report an empirical study of the failed patches generated by seven top LLM code agents. We collected 114 issues from the SWE-bench Lite dataset that remained unresolved across the agents. The seven agents produced a total of 769 failed patches for those issues, which we checked with a combination of GPT-4o and manual analysis. We present a taxonomy of the failure reasons across the patches. The taxonomy contains six categories, with several sub-categories under each category. For example, a frequently observed category is the inability of an LLM to correctly infer/produce the appropriate variable type in the produced patch. 
As a first step towards addressing such type-related errors, we designed \tool~ (Patch Agent). \tool~ utilizes program analysis techniques like CFG creation and exploration to infer the type information of a patch. \tool~ does this by applying repository-level static code analysis techniques. Then, \tool~ refines the inferred type by further utilizing an LLM-based inference technique. We tested \tool~ on all 127 type-related failed patches from the top three agents in our study. {\tool} could fix 29 of the 127 failed patches.

\end{abstract}



 \author{Haoran Xue}
 \email{hrx00@yorku.ca}
 \affiliation{%
   \institution{York University}
   \streetaddress{4700 Keele St.}
   \city{North York}
   \state{Ontario}
   \country{Canada}
   \postcode{M3J 1P3}
 }
  \author{Gias Uddin}
 \email{guddin@yorku.ca}
 \affiliation{%
   \institution{York University}
   \streetaddress{4700 Keele St.}
   \city{North York}
   \state{Ontario}
   \country{Canada}
   \postcode{M3J 1P3}
 }
  \author{Song Wang}
 \email{wangsong@yorku.ca}
 \affiliation{%
   \institution{York University}
   \streetaddress{4700 Keele St.}
   \city{North York}
   \state{Ontario}
   \country{Canada}
   \postcode{M3J 1P3}
 }


\setcopyright{none} 
\settopmatter{printacmref=false} 
\renewcommand\footnotetextcopyrightpermission[1]{}
\maketitle

\section{Introduction}
\label{sec:intro}

Software issue resolution requires the creation of a correct patch \cite{6698918, 10.1093/comjnl/bxv114}. As software complexity grows, the increasing volume and diversity of reported issues challenge traditional manual resolution, which is time-intensive and error-prone. This has driven interest in LLM-powered agents for automating issue resolution~\cite{he2024llmbasedmultiagentsystemssoftware,jin2024llmsllmbasedagentssoftware, xi2023risepotentiallargelanguage,10449667,zhang2024autocoderoverautonomousprogramimprovement,xia2024agentlessdemystifyingllmbasedsoftware,gauthier2024aider,getappmap2024specification}. 
Despite their potential, existing agent-based methods often fail to resolve a significant portion of issues effectively~\cite{10449667, 10.1145/3695988}. However, little is known about the underlying reasons for these failures. \textit{What are the root causes of unresolved issues of agents?} 
In this paper, we conduct an empirical study to investigate the underlying reasons behind failed patches generated by LLM-based agents for issue resolution.  
We seek to identify and understand the root causes of these failures, provide critical insights into the shortcomings of existing approaches, and inform the possible directions to enhance LLM-based agents for issue resolution.

Our study is based on the SWE-Bench Lite dataset~\cite{jimenez2023swe}, which provides a comprehensive benchmark for evaluating agent-based issue resolution approaches due to its diverse projects, issue types, and real-world complexity. We systematically collected and analyzed data from seven state-of-the-art models available on the SWE-bench Lite leaderboard and the Moatless EvalTool platform \cite{aorwall2024swebenchdocker} including AutoCodeRover(GPT-4o) \cite{zhang2024autocoderoverautonomousprogramimprovement}, Agentless (GPT-4o) \cite{xia2024agentlessdemystifyingllmbasedsoftware}, Aider (GPT4o \& Claude 3 Opus) \cite{gauthier2024aider}, AppMap Navie (GPT-4o) \cite{getappmap2024specification}, SWE-Agent (Claude 3 Opus) \cite{yang2024sweagentagentcomputerinterfacesenable} and Moatless Tools (GPT-4o) \& (Claude 3.5 Sonnet) \cite{moatless-tools}.

We manually collected 114 issues that remained unresolved across all seven models within the SWE-bench Lite dataset. To systematically analyze these failures, we employed a combination of an LLM-based framework and manual analysis to examine test failures and identify the underlying root causes of failures.
Based on this analysis, we developed a comprehensive taxonomy that categorizes six distinct failure patterns with sub-categories under each, providing a structured understanding of the key challenges faced by agent-based issue resolution methods. 

A frequently observed category in our taxonomy is the inability of issue resolution agents to correctly infer or produce the type of variable in the generated patch. Building on this finding, we propose {\tool} (Patch Agent) to improve the performance of the 
LLM-based issue resolution agents 
through targeted static analysis and type inference. {\tool} utilizes Abstract Syntactic Tree and program analysis techniques like Control Flow Graph (CFG) creation and exploration to infer type information. It also utilizes Large Language Models (LLMs) to refine inferred type information. 

We evaluated {\tool} on all 127 failed patches from the top three performing agents in our study (i.e., Agentless \cite{xia2024agentlessdemystifyingllmbasedsoftware}, Aider\cite{gauthier2024aider}, and AutoCodeRover \cite{zhang2024autocoderoverautonomousprogramimprovement}). Our evaluation demonstrates that {\tool} successfully resolved 29 previously failed patches, achieving a 22.83\% improvement rate on type-related issues. The most substantial enhancements were observed with Aider (GPT-4o \& Claude 3 Opus), where {\tool} fixed 11 previously unresolved patches, showing a 13.75\% relative improvement in performance. When applied to unresolved patches from Agentless (GPT-4o), {\tool}  successfully improved 10 patches, resulting in a 12.20\% relative improvement; and when applied to unresolved patches from AutoCodeRover (GPT4-o), {\tool} improved 8 patches, achieving 8.51\% relative improvement. These results highlight \tool's effectiveness in addressing a substantial portion of the type-related limitations in current agent-based issue resolution approaches.

\section{Empirical Study of Failed Patches from LLM-based Issue Fixing Agents}
\label{sec:empirical}
\begin{figure}[h]
    \centering    \includegraphics[width=.95\columnwidth]{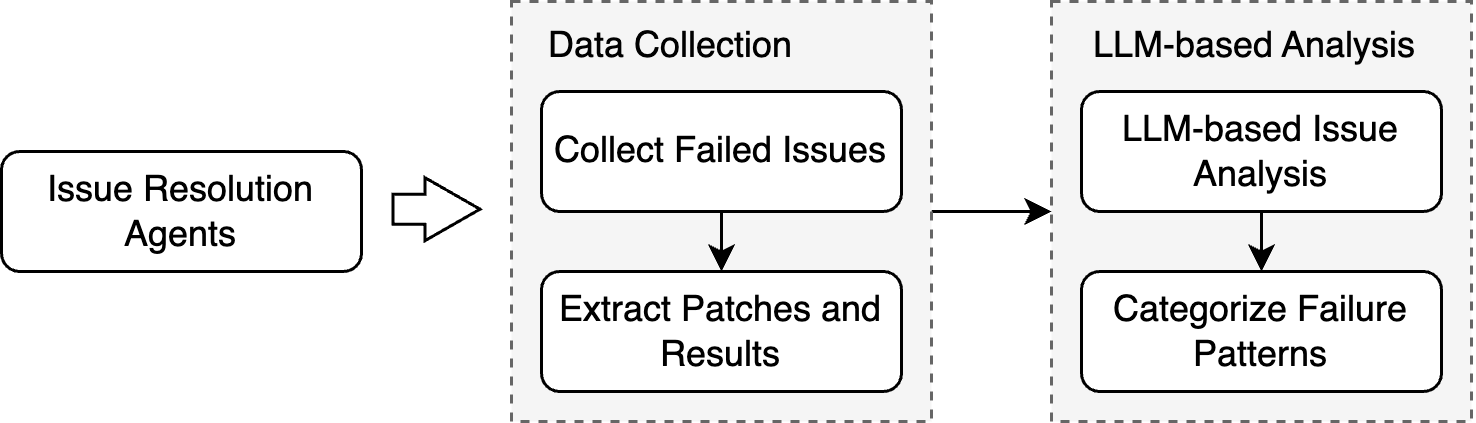} 
    \caption{Analysis steps for failed patches from LLM agents} 
    \label{fig: workflow_overview}
\end{figure}




\subsection{Studied LLM Code Agents}



We selected models based on their public and open-source availability on the SWE-bench lite leaderboard, which we can access their evaluation results and logs from their Github repositories or the Moatless Evaltool platform \cite{aorwall2024swebenchdocker}. Moatless is a validated evaluation platform that streamlines the assessment process for the SWE-bench datasets. Based on these requirements, we selected top seven models during the start time of our study (October 2024): 

\begin{enumerate}[leftmargin=10pt, itemsep=0pt]
    \item \textbf{AutoCodeRover (Resolution Rate: 30.7\%) } \cite{zhang2024autocoderoverautonomousprogramimprovement} integrates GPT4o \cite{openai2024gpt4ocard} with Abstract Syntax Tree (AST) representations, Spectrum-based Fault Localization (SBFL) \cite{52eb76f0306b479ba62297b00c6f7517}, and iterative patch refinement.

\item \textbf{Agentless (Resolution Rate: 27.3\%)} \cite{xia2024agentlessdemystifyingllmbasedsoftware} utilizes a seven-phase repair process with GPT4o.  

\item \textbf{Moatless Tools 1 (Resolution Rate: 26.7\%)} \cite{moatless-tools} is an integration of the Moatless framework with Claude 3.5 Sonnet \cite{anthropic2024claudea}.

\item \textbf{Aider (Resolution Rate: 26.3\%)} \cite{gauthier2024aider} leverages repository mapping and specializededit formats with GPT-4o and Claude 3 Opus. 

\item \textbf{Moatless Tools 2 (Resolution Rate: 24.7\%)} \cite{moatless-tools} leverages GPT4o's reasoning and code synthesis capabilities.

\item \textbf{AppMap Navie (Resolution Rate: 21.7\%)} \cite{getappmap2024specification} employs a deterministic seven-phase approach in GPT-4o with client-side context lookup.

\item \textbf{SWE-Agent (Resolution Rate: 11.7\%)} \cite{yang2024sweagentagentcomputerinterfacesenable} uses an agent-computer interface for repository interaction with Claude 3 Opus.

\end{enumerate}


\subsection{Data Collection and Extraction}


Our study specifically focuses on the SWE-bench Lite dataset, which is a streamlined version of the original SWE-bench benchmark. SWE-bench is a benchmark designed to evaluate the capabilities of automated bug-fixing and program repair agents in handling real-world software development issues \cite{jimenez2023swe}.

Our data collection process focused on unresolved issues common across the seven models we presented above within the SWE-bench Lite dataset. We manually collected 114 issues that remained unresolved across all seven models in our study. Since each issue corresponds to an individual case file for each model, our dataset consists of 798 case files in total. Each case file contained structured information, including the instance ID, issue type, problem description, gold patch (the correct fix accepted from the original GitHub pull requests), model-generated patch, and test output logs obtained from the Meatless platform. This systematic approach enabled us to perform a detailed comparative analysis across models while maintaining consistency in our evaluation criteria. 

\subsection{LLM-based Failed Patch Analysis}
To streamline our analysis of this extensive dataset, we leveraged an LLM-based framework to generate detailed reports, facilitating our manual evaluation. 
Our pipeline employs GPT-4o to examine each failed patch by considering multiple key aspects, including: 
(1) Test analysis - examines test failures to determine their root causes, identify the specific parts of the code being tested, and compare test behavior between the correct fix (gold patch) and the model-generated patch; 
(2) Patch comparison - analyze both syntactic and semantic differences between the gold patch and model-generated patch, identifying key differences; 
(3) Problem classification - analyze the bug type (e.g., logic error, API misuse), assess the domain knowledge required for resolution, and analyze the dependencies and contextual factors necessary for understanding the issue; 
(4) Model performance - investigate why the model-generated patch failed, identify recurring patterns in the model's approach, and assess whether the model correctly understood the underlying problem; 
(5) Repair strategy evaluation - compare the repair strategies used in the gold and model-generated patches, pinpoint missing knowledge or contextual understanding, and outline the reasoning steps required to arrive at a correct solution.

Through this comprehensive analytical framework, we processed a collection of failed patches, extracting essential information such as instance IDs, issue types, descriptions, gold patches, model-generated patches, and test outputs. We successfully processed 769 cases, with 29 cases excluded due to content length limitations in the LLM. The LLM-generated analysis results were stored in a structured JSON file, ensuring traceability and consistency for subsequent manual evaluation. 
 
\section{Taxonomy of Failed Patches}
\begin{figure*}[t]
    \centering \includegraphics[width=1\textwidth]{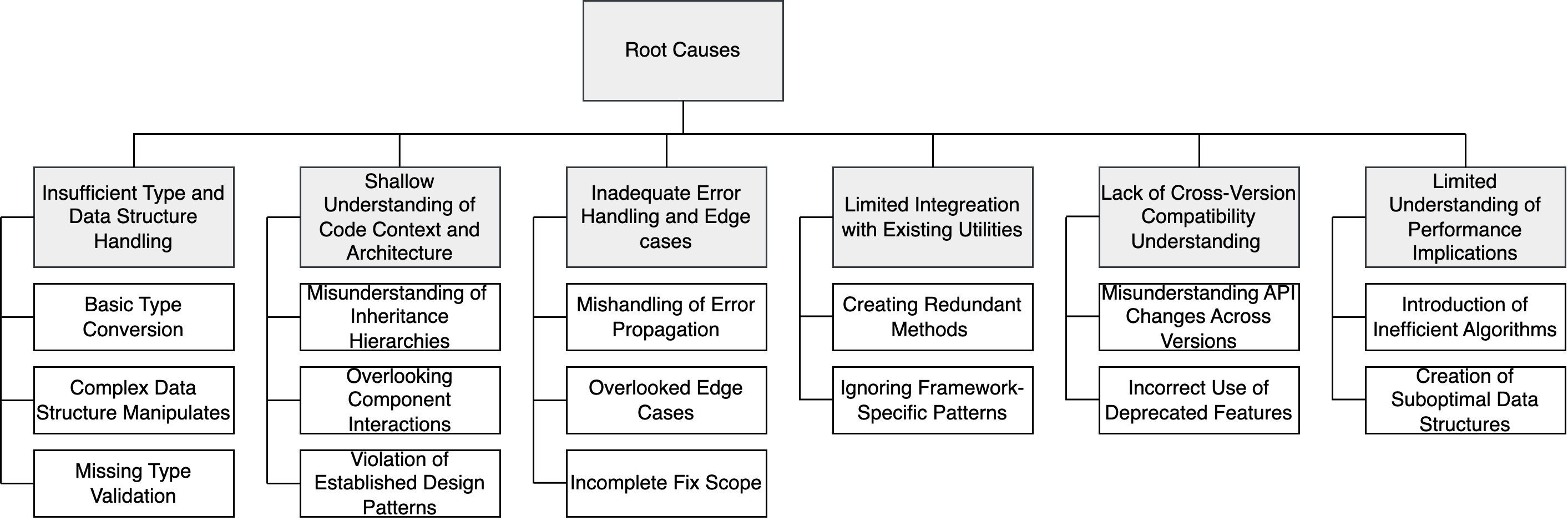} 
    \caption{Taxonomy of Failed Patches in Issue Resolution Agents}
    \label{fig:taxonomy}
\end{figure*}

Building on the LLM-generated patch analysis reports, we conducted a detailed manual review to identify recurring failure patterns. 
This helped us gain deeper insights into the fundamental limitations of current automated patch generation approaches and the common challenges faced by different models. By categorizing these failure patterns, we established a taxonomy of common failed patches shown in Fig.~\ref{fig:taxonomy}. 
Details of the six failure patterns are categorized as follows. 

 \subsection{Insufficient type and data structure handling} 
 This category represents cases where the issue resolution agents fail to properly handle data types and their transformations, including basic type conversions (i.e., a string to an integer) and complex data structure manipulations (i.e., numpy array to pandas dataframe). The key aspect is the LLM's limited ability to understand and manage how data types interact and transform within the system. This includes cases of missing type validation, incomplete conversions, or improperly managed data structures. These issues often manifest in libraries, including NumPy, pandas, scikit-learn, and django, where type conversions and data structure manipulations are frequent. In the issue scikit-learn\_\_scikit-learn-25638,  as illustrated in Table.~\ref{tab:patch-comparison-25638}, the issue involves handling pandas nullable dtypes (\texttt{int64, float64, and bool}) which, when converted to NumPy arrays, become \texttt{object} dtype, leading to be treated as mixed types which causes a \texttt{ValueError}. As shown in the Table, the gold patch (left column) addressed this by leveraging scikit-learn's built-in \texttt{check\_array} function, creating a reusable configuration dictionary with appropriate parameters for consistent application throughout the codebase. By replacing direct numpy conversions with \texttt{check\_array(y, dtype=None, ** check\_y\_kwargs)} and implementing proper error handling with specific checks, the gold patch provides a robust and maintainable solution that integrates seamlessly with the library's existing infrastructure. In contrast, in the Autocoderover(GPT4o)'s generated patch, it covered pandas nullable types to standard NumPy dtypes by applying \texttt{astype()}, replacing \texttt{``Int64''}, \texttt{``Float64''}, and \texttt{``boolean''} with \texttt{``int64''}, \texttt{``float64''}, and \texttt{``bool''}. 
 However, this approach hardcoded type replacements, which fails to generalize across other potential pandas extension types and potentially altering behaviors in edge cases. Moatless Tool 2 (Claude 3.5 Sonnet) introduced an explicit check for \texttt{hasattr(y, 'dtype')} and manually converted nullable pandas types using \texttt{to\_numpy()}. While this addressed dtype conversion at a basic level, it lacked structured type handling and failed to integrate with scikit-learn's existing processing functions like \texttt{check\_array}, which led to an inconsistent fix. 

\begin{table*}[t]
  \centering
  \footnotesize
  \setlength{\tabcolsep}{4pt}
   \caption{\small{Model Limitations in Data Type Handling - Comparing Gold Patch vs. Model-Generated Patches (scikit-learn\_\_scikit-learn-25638). Green texts indicate the correct implementation in the gold patch, while red text highlights the flawed approaches in model-generated patches.}}
   \vspace{-0.1in}
  \begin{tabular}{|p{0.33\textwidth}|p{0.33\textwidth}|p{0.33\textwidth}|}
    \hline
    \textbf{Gold Patch} & \textbf{AutoCodeRover+GPT4o Patch} & \textbf{Moatless+Claude 3.5 Sonnet Patch} \\
    \hline
    \begin{minipage}[t]{0.32\textwidth}
      \begin{Verbatim}[commandchars=\\\{\}]
#Correct fix: first conversion attempt with
 automatic dtype detection
+        \textcolor{green}{y = check_array(y, dtype=None,} 
+           \textcolor{green}{ **check_y_kwargs)}
# Fallback conversion with explicit object dtype
+        \textcolor{green}{y = check_array(y, dtype=object,} 
+           \textcolor{green}{**check_y_kwargs)}
      \end{Verbatim}
    \end{minipage}
    & 
    \begin{minipage}[t]{0.32\textwidth}
      \begin{Verbatim}[commandchars=\\\{\}]
# Wrong: Hardcoded type replacement
+ \textcolor{red}{ys = [y.astype(y.dtype.name.replace(}
+     \textcolor{red}{"Int64", "int64")}
+    \textcolor{red}{.replace("Float64", "float64")}
+    \textcolor{red}{.replace("boolean", "bool"))}
+    \textcolor{red}{if isinstance(y, pd.Series)}
+    \textcolor{red}{else y for y in ys]}
      \end{Verbatim}
    \end{minipage}
    &
    \begin{minipage}[t]{0.32\textwidth}
      \begin{Verbatim}[commandchars=\\\{\}]
# Wrong: Limited type checking
+    \textcolor{red}{if hasattr(y, 'dtype') and y.dtype.}
+       \textcolor{red}{name in ['Int64', 'Float64',}
+                \textcolor{red}{'boolean']:}
+       \textcolor{red}{ y = y.to_numpy()}
      \end{Verbatim}
    \end{minipage} \\
    \hline
  \end{tabular}
 
  \label{tab:patch-comparison-25638}
\end{table*}


\begin{table*}[t]
  \centering
  \footnotesize
  \setlength{\tabcolsep}{4pt}
    \caption{\small{Model Limitations in Code Context Understanding - Comparing Gold Patch vs. Model-Generated Patches (sympy\_\_sympy-18698)}}
    \vspace{-0.1in}
  \begin{tabular}{|p{0.31\textwidth}|p{0.36\textwidth}|p{0.32\textwidth}|}
    \hline
    \textbf{Gold Patch} & \textbf{Agentless+GPT4o Patch} & \textbf{AutoCodeRover +GPT4o Patch} \\
    \hline
    \begin{minipage}[t]{0.32\textwidth}
      \begin{Verbatim}[commandchars=\\\{\}]
#Correct fix: correctly merges factors with
the same multiplicity 
+   \textcolor{green}{if method == 'sqf':}
+       \textcolor{green}{factors = [(reduce(mul, (f for f, _ }
+          \textcolor{green}{in factors if _ == k)), k)}
+           \textcolor{green}{for k in set(i for _, i in} 
            \textcolor{green}{factors)]}
    return coeff, factors
      \end{Verbatim}
    \end{minipage}
    & 
    \begin{minipage}[t]{0.35\textwidth}
      \begin{Verbatim}[commandchars=\\\{\}]
#Wrong: attempts to merge but uses incorrect
approach: 
uses loop-based approach that concatenates
factors rather than properly combining them
+    \textcolor{red}{combined_result = []}
+    \textcolor{red}{for factor, multiplicity in result:}
+        \textcolor{red}{found = False}
+        \textcolor{red}{for i, (existing_factor,existing_multiplicity}
          \textcolor{red}{in enumerate(combined_result):}
+            \textcolor{red}{if existing_multiplicity == multiplicity:}
...       
      \end{Verbatim}
    \end{minipage}
    &
    \begin{minipage}[t]{0.32\textwidth}
      \begin{Verbatim}[commandchars=\\\{\}]
# Wrong: did the opposite of what's needed that
splits factors with higher multiplicities
+    \textcolor{red}{final_result = []}
+    \textcolor{red}{for factor, multiplicity in result:}
+        \textcolor{red}{if multiplicity > 1:}
+          \textcolor{red}{for \_ in range(multiplicity):}
+             \textcolor{red}{final\_result.append(}  
+             \textcolor{red}{(factor, 1)}
+        \textcolor{red}{else:}
+            \textcolor{red}{final_result.append((factor,}
             \textcolor{red}{multiplicity))}
      \end{Verbatim}
    \end{minipage} \\
    \hline
  \end{tabular}

  \label{tab:patch-comparison-18698}
\end{table*}

 {\centering
\begin{tcolorbox}
[width=1\columnwidth, boxrule=0.5pt, colback=gray!10, arc=4pt,
                  left=6pt, right=6pt, top=6pt, bottom=6pt, boxsep=0pt]
    \addtocounter{o}{1}\textbf{Observation \theo.} 
    Current issue resolution agents struggle with proper data type transformations, particularly when converting between complex data structures where type information must be carefully preserved.

\end{tcolorbox}
}

\subsection{Shallow understanding of code context and architecture} 
 
 This category encompasses cases where resolution agents fail to comprehend how components interact within a broader system architecture. It represents a fundamental limitation in understanding software beyond local functions or methods to grasp broader design patterns, architectural principles, and component relationships. The key aspect is the model's inability to reason about how localized changes impact the system as a whole. This includes cases where inheritance hierarchies are misunderstood, component interactions are overlooked, or established design patterns are violated. Table \ref{tab:patch-comparison-18698} illustrates this limitation with sympy\_\_sympy-18698, where the \texttt{sqf\_list} function fails to consistently handle factors with the same multiplicity. The gold patch correctly resolved the issue by ensuring the factors with the same multiplicity were merged into a single product, aligning the behavior of \texttt{sqf\_list} with the expected mathematical output. However, the model-generated patches failed to fully grasp the broader mathematical context. The patch generated by Agentless(GPT4o) attempts to combine factors with the same multiplicity, but it uses a loop-based approach inside \texttt{dup\_sqf\_list()}. The patch incorrectly handled factor merging that it altered the output structure instead of aligning with the expected mathematical combination of factors. AutoCodeRover's patch misunderstood the goal of the fix, it split factors with higher multiplicities into separate occurrences of multiplicity 1 instead of merging. 

 {\centering
\begin{tcolorbox}
[width=1\columnwidth, boxrule=0.5pt, colback=gray!10, arc=4pt,
                  left=6pt, right=6pt, top=6pt, bottom=6pt, boxsep=0pt]
    \addtocounter{o}{1}\textbf{Observation \theo.} 
    Current issue resolution agents often fail to grasp the broader software architecture, focusing on localized fixes without understanding component interactions, inheritance hierarchies, or established design patterns.

\end{tcolorbox}
}
    
 \subsection{Inadequate error handling and edge cases}This category represents instances where issue resolution agents implement simplistic solutions that address common cases but fail to handle errors, exceptions, or edge conditions comprehensively. This includes cases where error propagation is mishandled, edge cases are overlooked, or validation is oversimplified. In the sympy matrix case (sympy\_\_sympy-12454), the function incorrectly assumed a square matrix structure, iterating over all rows without ensuring that enough columns are available for valid indexing. This caused an IndexError when applied to rectangular (tall) metrics, as it attempted to access elements beyond the matrix's valid column range. The gold patch effectively resolved this by modifying both functions to ensure that column indices never exceeded the actual number of columns, preventing out-of-bounds errors while preserving correct behaviors for both square and rectangular matrices. In contrast, for the model generated patches, for instance, the SWE-Agent (Claude 3 Opus) and the Agentless (GPT4o), while correctly modifying \texttt{is\_upper()}, failed to apply the same fix to \texttt{\_eval\_is\_upper\_hessenberg()}, which indicates that a lack of comprehensive fix scope. 
 
 {\centering
\begin{tcolorbox}
[width=1\columnwidth, boxrule=0.5pt, colback=gray!10, arc=4pt,
                  left=6pt, right=6pt, top=6pt, bottom=6pt, boxsep=0pt]
    \addtocounter{o}{1}\textbf{Observation \theo.}
    Model-generated patches frequently implement simplistic solutions that handle common cases but overlook comprehensive error handling, exception management, and edge conditions.

\end{tcolorbox}
}
    
 \subsection{Limited integration with existing utilities} This category identifies cases where issue resolution agents create new functionality rather than leveraging existing utilities, helper methods, or established patterns within the codebase. The fundamental limitation is the model's inability to recognize and build upon the framework's existing abstractions and utility functions. This includes cases where redundant methods are created, existing utilities are overlooked, or framework-specific patterns are ignored. In sphinx\_\_sphinx-8273, the issue occurred because the generated man pages were replaced in a single flat directory, which did not conform to the Unix MANPATH requirement that expects man pages to be organized within section-specific directories. The gold patch correctly addressed this by introducing a configurable option \texttt{man\_make\_section\_directory}, ensuring flexibility and backward compatibility. It used the existing \texttt{ensuredir()} utility for directory creation, ensuring integration with the framework's established patterns and maintaining consistency. In contrast, the patches generated by Aider (GPT4o \& Claude 3 Opus) and by AppMap Naive (GPT4o) both failed due to limited integration with existing utilities and incomplete feature implementation. The patch generated by Aider (GPT4o) used Python's native \texttt{makedirs()} instead of existing \texttt{ensuredir()}, resulting in redundant logic and inconsistent framework integration. Similarly, the patch generated by AppMap Naive (GPT4o) manually manipulated strings to handle directory creation, again overlooking the utility already available in the framework.
 
 {\centering
\begin{tcolorbox}
[width=1\columnwidth, boxrule=0.5pt, colback=gray!10, arc=4pt,
                  left=6pt, right=6pt, top=6pt, bottom=6pt, boxsep=0pt]
    \addtocounter{o}{1}\textbf{Observation \theo.}
    Current issue resolution agents tend to create redundant functionality rather than leveraging existing utilities, helper methods, or established patterns within the codebase, which may lead to inconsistent framework integration.

\end{tcolorbox}
}
 \subsection{Lack of cross-version compatibility understanding} 
 
 This category encompasses cases where issue resolution agents fail to account for compatibility across different versions of libraries, frameworks, or language features. The key limitation is the model's inability to reason about how software evolves over time and how changes affect backward and forward compatibility. This includes cases where API changes are misunderstood, deprecated features are incorrectly used, or version-specific behaviors are overlooked. In a Matplotlib case (matplotlib\_\_matplotlib-24970), the issue involved NumPy API changes that modified how out-of-range integers are handled when converting to uint8. Matplotlib's colormap lookup previously relied on implicit overflow behavior, which was deprecated in NumPy 1.24. The gold patch handled the issue by ensuring that values outside the allowed range for colormap indices were adjusted to stay within the valid limits before being used. In simpler terms, it prevented numbers that were too big or too small from causing errors by setting a clear boundary for acceptable values. This fix followed NumPy 1.24's updated rules for handling numbers and ensured that Matplotlib would work correctly with both older and newer versions of NumPy without errors. However, some model-generated patches attempted various fixes but exhibited key shortcomings. For example, Aider (GPT4o \& Claude 3 Opus)'s patch relied on multiple redundant\texttt{np.clip(x, a\_min, a\_max)} calls, unnecessarily clamping values multiple times without addressing the core issue efficiently. The patch generated by Moatless Tool (GPT4o)'s attempted a minimal fix using \texttt{np.clip(x, a\_min, a\_max)}, but it removed explicit handling for over-range and under-range cases, potentially altering expected behavior. These issues illustrate a common challenge in automated repair systems, which is a lack of cross-version compatibility understanding. The patches failed to account for how numpy's integer type handling evolved, leading to fixes that either missed necessary constraints or introduced unintended behavior. 
 
 {\centering
\begin{tcolorbox}
[width=1\columnwidth, boxrule=0.5pt, colback=gray!10, arc=4pt,
                  left=6pt, right=6pt, top=6pt, bottom=6pt, boxsep=0pt]
    \addtocounter{o}{1}\textbf{Observation \theo.} Current issue resolution agents struggle with cross-version compatibility issues, failing to account for API changes, deprecated features, or version-specific behavior across different library versions.

\end{tcolorbox}
}
 \subsection{Limited understanding of performance implications} This category represents instances where issue resolution agents implement functionally correct patches but fail to consider their impact on system performance, efficiency, or resource utilization. The key limitation is the model's inability to reason about computational complexity, resource usage patterns, or optimization opportunities beyond achieving basic functional correctness. This includes cases where insufficient algorithms are introduced, redundant computations are performed, or suboptimal data structures are created. In sympy's sympy\_\_sympy-19542, illustrate this limitation with \texttt{sympy\_\_factortools}, where the \texttt{dup\_zz\_mignotte\_bound} function initially used an outdated Mignotte bound that resulted in significant overestimation, leafing to inefficient computations. The gold patch correctly resolved the issue by replacing it with the Knuth-Cohen bound, applying binomial coefficients, Euclidean norms, and specific handling for irreducible polynomials, which significantly optimized the bound estimation. However, the model-generated patches failed to grasp the mathematical complexity and performance implications. The patch by Agentless retained the original flawed formula with minimal adjustments, failing to improve performance. AutoCodeRover's patch introduced an incorrect simplified formula that omitted key mathematical components like binomial coefficients and Euclidean norms, leading to inaccurate estimations. Both patches demonstrate a limited understanding of algorithmic optimization, resulting in suboptimal solutions.

 {\centering
\begin{tcolorbox}
[width=1\columnwidth, boxrule=0.5pt, colback=gray!10, arc=4pt,
                  left=6pt, right=6pt, top=6pt, bottom=6pt, boxsep=0pt]
    \addtocounter{o}{1}\textbf{Observation \theo.} While implementing functional correctness, issue resolution agents generated patches often ignore the performance implications, implementing solutions with inefficient algorithms and redundant computations.

\end{tcolorbox}
}

\subsection{Distribution of analyzed patches across categories}
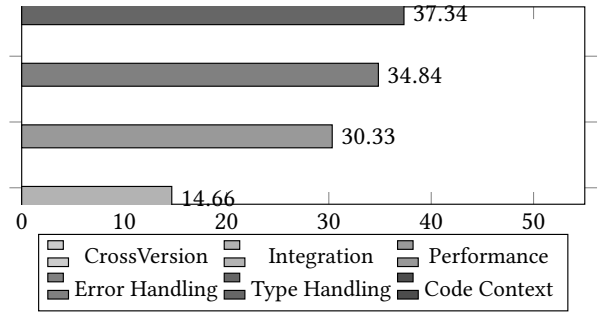
\begin{figure}[t]
    \centering
    \begin{adjustbox}{width=0.95\linewidth}
    \begin{tikzpicture}
        \begin{axis}[
            width=\linewidth,
            height=4cm,
            xbar,
            bar width=8pt,
            symbolic y coords={ CrossVersion, Integration, Performance, Error Handling, Type Handling, Code Context},
            symbolic y coords,
            xmin=0,
            xmax=55,
            nodes near coords,
            nodes near coords align={horizontal},
            xtick={0,10,20,30,40,50},
            legend style={at={(0.5,-0.15)}, anchor=north, legend columns=3, font=\small},
            xlabel={Percentage (\%)},
        ]
        \addplot[fill=black!20] coordinates {(8.52,CrossVersion)};
        \addplot[fill=black!30] coordinates {(14.66,Integration)};
        \addplot[fill=black!40] coordinates {(30.33,Performance)};
        \addplot[fill=black!50] coordinates {(34.84,Error Handling)};
        \addplot[fill=black!60] coordinates {(37.34,Type Handling)};
        \addplot[fill=black!70] coordinates {(47.74,Code Context)};
        \vspace{2cm}
        \legend{CrossVersion, Integration, Performance, Error Handling, Type Handling, Code Context}
        \end{axis}
    \end{tikzpicture}
    \end{adjustbox}
    \caption{Distribution of Analyzed Patches Across Categories}
    \label{fig:patches across caterogies}
\end{figure}

For each category, we computed statistics on pattern frequencies. As shown in Fig. \ref{fig:patches across caterogies}, among the patches generatedby the seven models, patches related to ``Insufficient handling of data types and structures'' (37.34\%) and ``Shallow understanding of code context and architecture'' (47.74\%) were the most common failure patterns across all models. ``Inadequate error handling'' (34.84\%) and ``Limited understanding of performance implications'' (30.33\%) also represented significant portions of the failed patches, while ``Limited integration with existing utilities'' (14.66\%) and ``Lack of cross-version compatibility understanding'' (8.52\%) were less common but still noteworthy. This distribution highlights that the primary challenges in agent-generated patches are mainly due to type handling and architectural understanding, which informed our decision to focus on type-related improvements in our tool development. 
Table. \ref{tab:llm-error-counts} presents the specific counts of failed patches exhibiting each failure pattern across the different models. 

\begin{table*}[t]
\centering
\caption{Distribution of failure patterns across agents}
\vspace{-0.1in}
\setlength{\tabcolsep}{4pt}
\begin{tabular}{@{}lccccccc@{}}
\toprule
\textbf{ $\downarrow$ Failure Patterns | Agents $\rightarrow$} 
 & \textbf{Agentless} & \textbf{Aider} & \textbf{AppMap} & \textbf{AutoC} & \textbf{Moatless 1} & \textbf{Moatless 2} & \textbf{SWEA} \\
 & \textbf{(GPT4o)} & \textbf{(GPT4o \& Claude 3)} & \textbf{Naive (GPT4o)} & \textbf{(GPT4o)} & \textbf{(GPT4o)} & \textbf{(Claude3.5)} & \textbf{(Claude3)} \\ 
\midrule
Type \& data structure & 17 & 14 & 12 & 13 & 14 & 18 & 10 \\
Code context & 34 & 41 & 40 & 37 & 41 & 35 & 25 \\
Inadequate error handling & 56 & 47 & 48 & 50 & 45 & 39 & 59 \\
Cross-version compatibility & 5 & 5 & 5 & 6 & 6 & 7 & 8 \\
Integration with utilities & 0 & 4 & 5 & 2 & 2 & 3 & 7 \\
Performance implications & 0 & 0 & 1 & 0 & 1 & 1 & 0 \\
\bottomrule
\end{tabular}
\label{tab:llm-error-counts}
\end{table*}

\section{\tool: Enhancing Issue Resolution Agents with Data Type Support}
\label{sec:approach}

Based on our empirical study, we found that `\texttt{Insufficient handling of data type and structure}''  is a significant limitation in existing LLM-based bug fixer agents. While not the most frequent failure pattern, it represents a particularly promising area for improvement. Unlike failures caused by a lack of architectural awareness, where a model may need to understand an entire system's design, errors related to data type handling often arise from specific, localized issues such as improper conversions, incorrect use of data structures, or missing type validation. Type-related failures often stem from specific, identifiable patterns that can be systematically addressed through targeted static analysis and inference techniques.

We developed a specialized agent called {\tool} (Patch Agent). This section presents the detailed design and implementation of our approach to enhance existing issue resolution agents by addressing their limitations in handling data type and structure issues.


\subsection{Overview and Objectives}
The primary objective of {\tool} is to enhance patches generated by LLM-based bug fixers by focusing on resolving data type and structure issues that the original systems struggle with. {\tool} functions as a post-processing layer that can be integrated with existing agent-based issue resolution systems. It takes as input a candidate patch from an issue resolution agent, analyzes it for potential type and structure issues, and produces a refined patch with improved type handling.

\begin{figure*}[ht!]
    \centering    \includegraphics[width=0.99\textwidth]{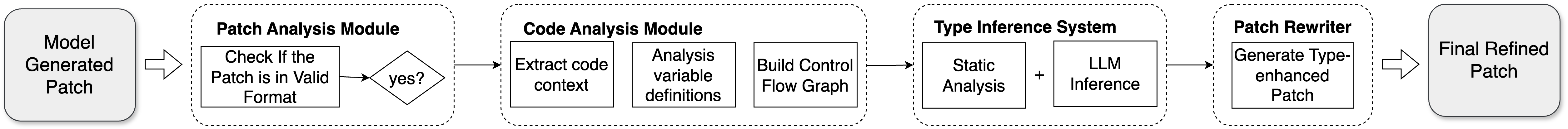} 
    \vspace{-0.1in}
    \caption{Overall workflow of {\tool}} 
    \label{fig: overall_workflow_pagent}
    \end{figure*}

Fig.\ref{fig: overall_workflow_pagent} shows the overview of {\tool},  which contains four main components, i.e., 
\textbf{1. Patch Analysis Module} (see Section~\ref{sec:4.2}): Examines the candidate patch for potential type and structure issues. \textbf{2. Code Analysis Module} (see Section~\ref{sec:4.3}): Performs static analysis of the code context relevant to the patch, extracting type information and code structure.
\textbf{3. Type Inference System} (see Section~\ref{sec:4.4}): Leverage both static analysis and LLM capabilities to infer appropriate types for variables in the context.
\textbf{4. Patch Rewriter} (see Section~\ref{sec:4.5}): Refines the candidate patch to address identified type and structure issues while preserving the intended functionality. 

\subsection{Patch Analysis Module}
\label{sec:4.2}

This module examines candidate patches to identify potential type and structure issues. It focuses on (1) validating the patch format: ensuring the patch adheres to standard unified diff format and can be applied correctly; and (2) Analyzing Code Syntax: validating the syntactic correctness of the code in the patch. To ensure patch integrity, we first validate the patch format through a multi-tiered verification process. Our system checks for the presence of standard unified diff markers, such as \texttt{diff --git, +++, ---, and @@}, and also verifies it contains proper file headers, correct formatted file paths, and syntactically valid hunk markers that follow the \texttt{@@ -start,count +start, count@@} pattern. This validation is essential because malformed patches would inevitably fail during the application, wasting computational resources and subsequent type analysis and refinement sites that could never succeed. (example showing correct header).

For syntactic correctness validation, we extract the added code lines (those prefixed with \texttt{+} but not \texttt{+++}) from each modified file block within the patch. These are then reassembled into coherent code blocks and parsed using Python's abstract syntax tree (AST). This approach allows us to verify that the proposed changes represent syntactically valid Python code. By detecting syntax errors early, we prevent cascading failures in altering pipeline stages and provide more actionable feedback for patch improvement.

We provide an example to show how {\tool} processes a django issue with Aider (GPT4o\&Claude 3 Opus) as provided in Fig. \ref{fig:pagent-exampel-14534}. When analyzing this issue, in the upper-left block of Fig.\ref{fig:pagent-exampel-14534}, {\tool} 
begins by ingesting the original patch generated by Aider (GPT4o \& Claude 3 Opus) Agent. We first check this patch with our Patch Analysis module, which is depicted in the upper right block, to see if the format of the patch is correctly applied through the Patch Analysis Module. The module validates the unified diff format, confirming the presence of proper headers (diff --git), file paths, and index identifiers) and correctly formatted hunk markers (@@ -227,7 +227,7 @@).

\subsection{Code Analysis Module}
\label{sec:4.3}

The code analysis module implements advanced static analysis techniques to extract meaningful type information and structural context from the entire code repositories, which contain the files that the patch is modifying. The primary purpose of this step is to overcome the limitation of localized patch examination that leads LLM-based bug fixers to fail with type-related issues. By analyzing the entire code repositories, not just the patch-modified ones, we capture crucial contextual information about how types and data structures are used throughout the project. 

We consider the dynamic of Python's typing system, which means that a variable might be assigned a pandas DataFrame in one module but transformed or accessed in another module touched by the patch. Therefore, we need a comprehensive understanding beyond the immediate scope of the patch. Without confirming this, even sophisticated LLMs will struggle to infer the appropriate type of handling. We first parse Python source files in the code repository into abstract syntax tree (AST) representations, then we traverse each AST representation to locate variable definitions, assignments, and usage patterns, particularly on variables that appear in the patch. For each identified variable, we apply heuristics to infer its type based on assignment patterns, for example, for assignments with literal values (e.g., x = 5), we directly infer the type based on the literal (int), for assignments with container constructors (e.g., x = list(), y = dict()), we infer the corresponding container type, for assignments with expressions, we track the expression type when possible. This AST-based approach provides a first-pass type inference that captures straightforward type definitions in the codebase, serving as a foundation for our other analyses. 

For each function affected by the patch, we build a control flow graph (CFG) to represent possible execution paths and how variables might be transformed. 
We then implement reaching definitions analysis over the control flow graph to track variable definitions and their potential values across different execution paths. The reaching definitions analysis operates through an iterative algorithm that propagates variable definition information across the CFG until a fixed point is reached. For each node in the CFG, it computes two key sets per node: incoming definitions from predecessors and outgoing definitions after execution. When a variable is redefined within a node, this new definition will "kill" previous definitions of the same variable in the reaching definitions. This process creates a comprehensive map of valid variable definitions at each program point, enabling precise type tracking across execution paths and revealing subtle type-related bugs that might only appear in specific execution scenarios. These analysis techniques allow us to build a comprehensive understanding of the code context surrounding the variables and expressions in the patch. This is essential for accurately inferring types and understanding complex interactions between different code components. 

In our django issue example, after validating the patch format, we passed this patch to the Code Analysis Module, shown in the middle-left block of Fig. \ref{fig:pagent-exampel-14534} where we systematically processed the django repository structure, analyzing over 300 Python files in the tests directory to understand the broader context. From the test paths like workspace\_comp/repo\_django\_\_django-14534/tests/
template\_tests, workspace\_comp/repo\_django\_\_django-14534/tests/
forms\_tests, and other test directories,  {\tool} builds a comprehensive understanding of how form widgets and their attributes are used throughout the codebase. When analyzing the AST of \texttt{boundfield.py}, PAGENT identified five critical variables for type analysis: `self', `self.data', ``self.data['name']'', ``self.data['index']'', and ``self.data['attrs']''. Through Control Flow Graph (CFG) construction, {\tool} tracked how these variables flowed through the \texttt{BoundWidget} class, particularly focusing on the property path from \texttt{BoundField.subwidgets} to \texttt{BoundWidget.id\_for\_label}, which revealed that 'attrs' carried widget IDs throughout the rendering process. 

\subsection{Type Inference System}
\label{sec:4.4}

The type inference system represents a crucial innovation in code maintenance by combining traditional static analysis with LLMs' capabilities. This hybrid approach emerged from the realization that conventional static analysis tools often struggle with complex Python codebase, where type information is implicit or spread across multiple files.

The system builds upon the foundation established by the initial code analysis phase, extending beyond basic static analysis to address the complexities of Python's dynamic typing, which is discussed in the Code Analysis module. While the code analysis module provides essential structural understanding through AST parsing, CFG construction, and variable tracking, the type inference system focuses specifically on deriving meaningful type information from this analysis. The system intelligently prioritizes code segments, emphasizing type-relevant patterns such as assignments, function signatures, and method invocations to create a distilled content representation that captures the semantic essence of variable usage. 

For variables where conventional static analysis proves insufficient, particularly those involving complex data structures, user-defined classes, or domain-specific types, we leverage large language models (LLMs) to interpret the contextual nuances. The LLM component examines variable usage patterns within their full execution context, mimicking how experienced developers intuitively understand types in dynamic languages by observing behavior rather than declarations. 

By integrating static analysis with LLM-powered inference, our system achieves a balance of precision and flexibility that neither approach could provide independently. This hybrid methodology allows us to generate appropriate type annotations even for complex codebases where type information is implicit or distributed across multiple modules. The final output consists of specific type suggestions for each variable identified in the patch, providing developers with semantically rich type annotations that enhance code clarity and enable more effective static analysis without sacrificing the expressive power of Python's dynamic typing system.

In our django issue example, the Type Inference System, illustrated in the lower-left block of Fig. \ref{fig:pagent-exampel-14534}, leveraged both static analysis and LLM capabilities to determine more accurate types for the variables. Through contextual pattern recognition,  {\tool} inferred that \texttt{self.data['name']} is a string, \texttt{self.data['index']} likely represents a positional identifier, and \texttt{self.data['attrs']} is a dictionary structure containing optional ID values. This typing information provides crucial context for understanding the variable relationships and expected return types.

\subsection{Patch Rewriter} 
\label{sec:4.5}

This module uses the inferred type information and code context from previous steps to produce type-enhanced patches. We prompt an LLM to write the patch with strict structures the prompt to emphasize the importance of maintaining proper diff format, including all headers, hunk markers, and line prefixes that are essential for proper patch application. Beyond basic syntax validation, we integrate mypy static type checking from Python to analyze the type annotations for consistency and correctness. In the end, we get the valid refined patches.

\begin{figure}[t]
    \centering \includegraphics[width=.5\textwidth]{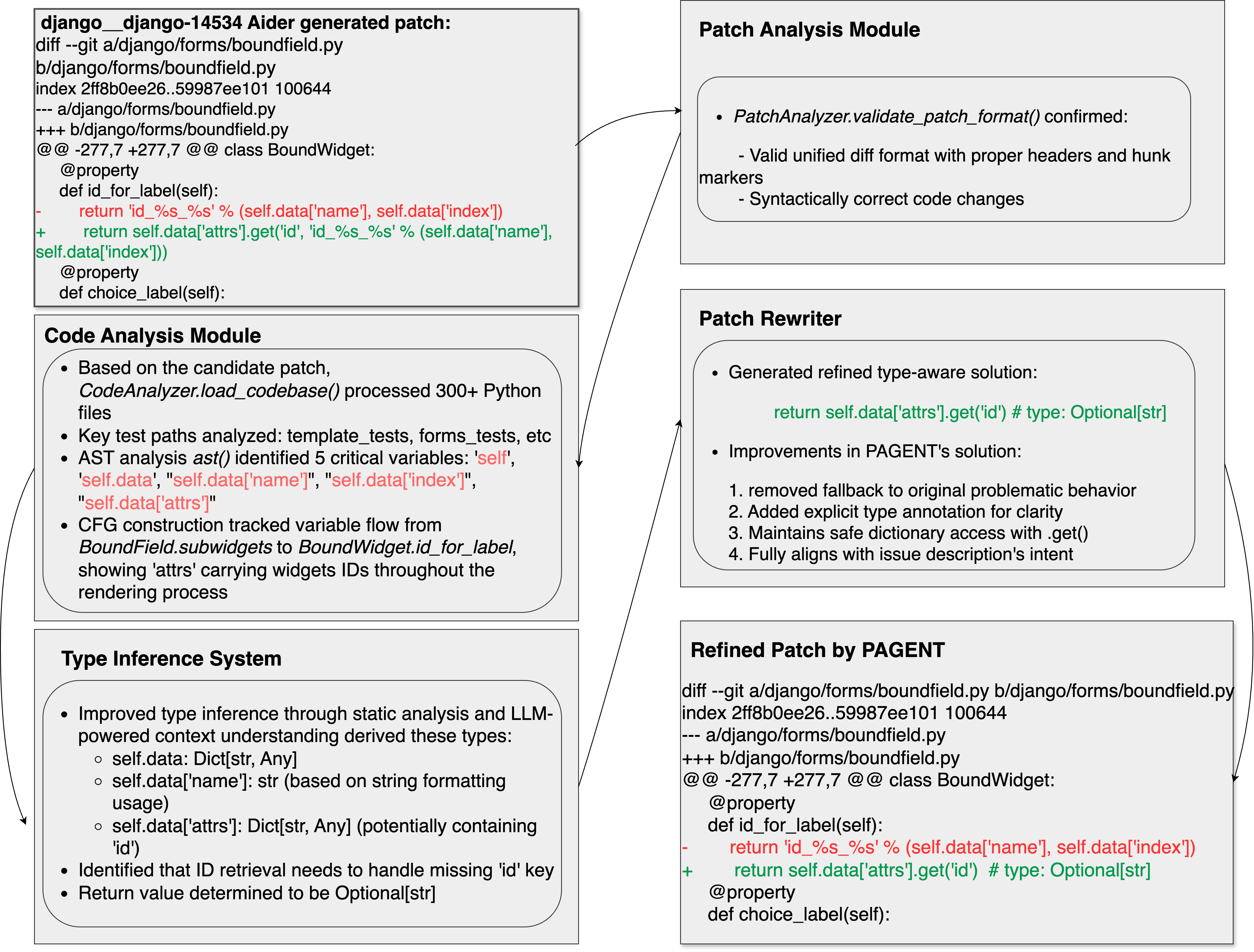} 
    \caption{An example of how {\tool} processes an issue of django\_\_django-14534 to refine it progressively}
    \label{fig:pagent-exampel-14534}
\end{figure}

For the django issue example, the Patch Rewriter presented in the middle-right block of Fig. \ref{fig:pagent-exampel-14534} used the type information gathered in previous steps to improve the original patch. With the knowledge that \texttt{self.data['attrs']} is a dictionary that may contain an 'id' key, which could be absent, the rewrite recognized that the return value of this function could be either a string or None. This led to the improved patch that replaced the original implementation with a simple return \texttt{self.data['attrs'].get('id')} alongside the appropriate type annotation \texttt{\# type: Optinal[str]}. In the end, the bottom-right block of Fig. \ref{fig:pagent-exampel-14534} shows the final refined patch produced by  \tool, clearly displaying how the original problematic line was replaced with a more robust, type-safe implementation.

\section{Effectiveness Assessment of \tool}
\label{sec:results}



We answer three research questions (RQs):

\begin{enumerate}[label=\textbf{RQ\arabic{*}.}]
\item {How effective is {\tool} in addressing previously unresolved type-related errors?} 

\item {How effective is {\tool} in improving the performance of existing models?}

\item {How effective is {\tool} in handling the sub-categories of type-related issues?}
\end{enumerate}

\subsection{Experimental Setup}
We established a comprehensive evaluation framework using the SWE-bench Lite dataset. Our evaluation focused on the failed patches that were identified in our Section \ref{sec:empirical} as exhibiting insufficient type and data structure handling problems. In total, there are 762 patches, representing approximately 37.34\% of all failed patch attempts across the seven resolution agents. 
We employed the Moatless online evaluation tool \cite{aorwall2024swebenchdocker} to evaluate the refined patches generated by our tool, which provides a streamlined and rigorous assessment process for evaluating the SWE-bench Lite dataset. This platform allows us to validate the code repository and run the project's test suite, ensuring that our evaluation maintains the high standards of the SWE-bench benchmark while enabling more efficient testing. 

\subsection{Overall Performance (RQ1)}

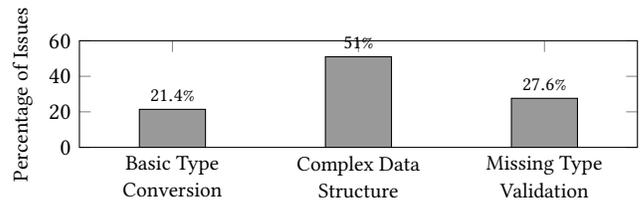
\begin{figure}[ht!]
    \centering
    \begin{tikzpicture}
        \begin{axis}[
            ybar,
            bar width=25pt,
            height=3cm,
            width=9cm,
            enlarge x limits=0.25,
            ylabel={Percentage of Issues},
            ylabel style={font=\small},
            symbolic x coords={Basic Type Conversion, Complex Data Structure, Missing Type Validation},
            xtick=data,
            ymin=0, ymax=60,
            nodes near coords={\pgfmathprintnumber\pgfplotspointmeta\%},
            every node near coord/.append style={font=\footnotesize},
            legend style={at={(0.5,1.05)},anchor=south,legend columns=-1},
            xticklabel style={font=\small},
            xtick pos=both,
            xtick align=inside,
            xticklabels={\shortstack{Basic Type\\Conversion}, \shortstack{Complex Data\\Structure}, \shortstack{Missing Type\\Validation}}
        ]
        \addplot[fill=black!40!white] coordinates {
            (Basic Type Conversion, 21.4) 
            (Complex Data Structure, 51.0) 
            (Missing Type Validation, 27.6) 
        };
        \end{axis}
    \end{tikzpicture}
    \vspace{-0.1in}
    \caption{Distribution of unfixed patches across subcategories}
    \label{fig:sub-cat_unfix}
\end{figure}

We applied {\tool} to the failed patches that exhibited type-related issues cross all seven agents.
PAGENT was able to fix patches for three agents: Agentless (GPT4o), AutoCodeRover (GPT4o), and Aider (GPT4o \& Claude 3 Opus). For these three agents, {\tool} successfully resolved 29 previously failed patches out of 127 type-related failures, achieving an improvement rate of 22.83\%. Fig.\ref{fig:sub-cat_unfix} provides valuable insights into the distribution of challenges that remain unresolved. 

We provide an example case, sphinx-doc\_\_sphinx-7738 with Agentless+GPT4o model to illustrate \tool's type understanding and contextual awareness, as detailed in Table. \ref{tab:PAGENT-example-7738}. In this case, {\tool} successfully addressed a critical type-related issue in the sphinx project. The original patch from Agentless + GPT4o attempted to fix a string handling issue in the \texttt{escape\_args\_and\_kwargs} method using a naive string comparison \texttt{and not name.endwith(r'\_')}. However, this approach misunderstood both Python's type system and the architectural design of the codebase.  It led to an AssertionError in the test \texttt{text\_underscore\_in\_attribute} as shown in the test results in Table.\ref{tab:PAGENT-example-7738}, where underscores in attribute names were incorrectly escaped, resulting in \texttt{arg\_} appearing as \texttt{arg\ \_} even when it should have remained unmodified based on configuration settings. {\tool} identified that this behavior should be controlled by configuration rather than hardcoded string manipulation. It proposed using \texttt{getattr(self.\_config, 'strip\_signature\_backslash', False)}, ensuring the escape behavior respects user preferences and aligns with the codebase's architectural patterns. Our test results confirmed that this refined approach effectively resolved the issue. 

\begin{table*}[t]
  \centering
  \footnotesize
  \setlength{\tabcolsep}{4pt}
   \caption{Comparison of patches generated by Agentless+GPT4o and {\tool} for the sphinx-doc/sphinx-7738 issue.} 
  \label{tab:PAGENT-example-7738}
  \begin{tabular}{|p{0.5\textwidth}|p{0.5\textwidth}|}
    \hline
    \textbf{Original Patch Generated by Agentless+GPT4o} & \textbf{Refined Patch Generated by PAGENT} \\
    \hline
    \begin{minipage}[t]{0.49\textwidth}
      \begin{Verbatim}[commandchars=\\\{\}]
@@ -318,7 +318,7 @@ class GoogleDocstring:
            return [line[min_indent:] for line in lines]
    def escape_args_and_kwargs(self, name: str) -> str:
-        \colorbox{pink}{if name.endswith('_'):}
\textcolor{red}{#Wrong fix: Uses naive string pattern matching that ignores the}
\textcolor{red}{configuration system} 
+        \colorbox{green!50!yellow}{if name.endswith('_') and not name.endswith(r'\_'):}
            name = name[:-1] + r'\_'
        if name[:2] == '**':
      \end{Verbatim}
    \end{minipage}
    & 
    \begin{minipage}[t]{0.49\textwidth}
      \begin{Verbatim}[commandchars=\\\{\}]
@@ -318,7 +318,7 @@ class GoogleDocstring:
            return [line[min_indent:] for line in lines]
    def escape_args_and_kwargs(self, name: str) -> str:
-        \colorbox{pink}{if name.endswith('_'):}
\textcolor{green}{#Correct Fix: Make underscore escaping optional and avoid hardcoding}
\textcolor{green}{the behavior}
+        \colorbox{green!50!yellow}{if name.endswith('_') and getattr(self._config,}
        \colorbox{green!50!yellow}{'strip_signature_backslash', False):}
            name = name[:-1] + r'\_'
        if name[:2] == '**':
      \end{Verbatim}
    \end{minipage} \\
    \hline
    \textbf{Test Results (Agentless+GPT4o)} & \textbf{Test Results (PAGENT)} \\
    \hline
    \begin{minipage}[t]{0.49\textwidth}
      \begin{Verbatim}[commandchars=\\\{\},fontsize=\scriptsize]
============================= FAILED TESTS ==============================
NumpyDocstringTest.test_underscore_in_attribute

Expected:
  :ivar arg_: some description
  :vartype arg_: type
Actual:
  :ivar arg\_: some description   (backslash added)
  :vartype arg\_: type   (backslash added)

\textcolor{red}{AssertionError:} expected and actual output do not match
      \end{Verbatim}
    \end{minipage}
    & 
    \begin{minipage}[t]{0.49\textwidth}
      \begin{Verbatim}[commandchars=\\\{\}]
=========================== test session summary =======================
collected 31 items
tests/test_ext_napoleon_docstring.py ..........................   [100%]

==========================short test summary info =======================
PASSED tests/test_ext_napoleon_docstring.py::NumpyDocstringTest::
test_underscore_in_attribute
PASSED tests/test_ext_napoleon_docstring.py::NumpyDocstringTest::
test_underscore_in_attribute_strip_signature_backslash
====================== 31 passed, 7 warnings in 0.38s ==================
      \end{Verbatim}
    \end{minipage} \\
    \hline
    
  \end{tabular}
 
\end{table*}


{\centering
\begin{tcolorbox}
[width=1\columnwidth, boxrule=0.5pt, colback=gray!10, arc=4pt,
                  left=6pt, right=6pt, top=6pt, bottom=6pt, boxsep=0pt]
    \textbf{RQ1 Summary.}
    {\tool} successfully resolved 29 out of 127 previously failed patches with type and data structure handling problems, achieving a 22.83\% improvement rate.

\end{tcolorbox}
}

\subsection{Model-Specific Improvements (RQ2)}
The application of {\tool} as a post-processing step to patches previously generated by LLM-based issue resolution agents demonstrates notable improvements across the evaluated models, with particularly notable enhancements for three of the seven evaluated models. Table \ref{tab:pagent_improvement} summarizes these results. The most substantial enhancements were observed when applied to patches from the Aider (GPT 4o \& Claude 3 Opus), where PAGENT successfully fixed 11 previously unresolved patches, thereby increasing Aider's effective success rate from 26.67\% to 30.33\%. This 13.75\% relative improvement demonstrates PAGENT's ability to identify and rectify type-related shortcomings in patches that otherwise contained promising repair strategies.

Similarly, when applied to previously unresolved patches from Agentless (GPT4o), PAGENT fixed 10 patches, raising the effective success rate from 27.33\% to 30.67\%, a 12.20\% relative improvement. For the AutoCodeRover (GPT4o), PAGENT successfully fixed 8 patches. Overall, {\tool} successfully fixed three common issues across the models.

\begin{table}
%
\caption{\small{Performance improvement with {\tool} integration}}
\vspace{-0.1in}
\label{tab:pagent_improvement}
\resizebox{\columnwidth}{!}{
\begin{tabular}{lcccc}
\toprule
\textbf{Issue Resolution Agents} & \textbf{Original} & \textbf{Post-PAGENT Rate} & \textbf{Impr.} & \textbf{Rel. Gain} \\
\midrule
\textbf{Aider} & 80/300 (26.67\%) & 91/300 (30.33\%) & \textcolor{red}{+11} & \textcolor{red}{+13.75\%} \\
\textbf{Agentless} & 82/300 (27.33\%) & 92/300 (30.67\%) & \textcolor{red}{+10} & \textcolor{red}{ +12.20\%} \\
\textbf{AutoCodeRover} & 94/300 (31.33\%) & 102/300 (34.00\%) & \textcolor{red}{+8} & \textcolor{red}{+8.51\%} \\
\textbf{Moatless Tool 1} & 35/300 (11.67\%) & 35/300 (11.67\%) & 0 & 0\% \\
\textbf{Moatless Tool 2} & 41/300 (13.67\%) & 41/300 (13.67\%) & 0 & 0\% \\
\textbf{SWE-Agent (Claude 3 Opus)} & 35/300 (11.67\%) & 35/300 (11.67\%) & 0 & 0\% \\
\textbf{AppMap Naive (GPT-4o)} & 44/300 (14.67\%) & 44/300 (14.67\%) & 0 & 0\% \\
\bottomrule
\end{tabular}%
}
\end{table}

{\centering
\begin{tcolorbox}
[width=1\columnwidth, boxrule=0.5pt, colback=gray!10, arc=4pt,
                  left=6pt, right=6pt, top=6pt, bottom=6pt, boxsep=0pt]
    \textbf{RQ2 Summary.}
    {\tool} demonstrated model-specific improvements when applied as a post-processing step. The consistent improvement across different models suggests that {\tool} effectively addresses common type-related issues regardless of the underlying model architecture.

\end{tcolorbox}
}
\subsection{Sub-Category Effectiveness (RQ3)}

\begin{figure}[t!]
    \centering
    \begin{tikzpicture}
        \begin{axis}[
            ybar,
            bar width=25pt,
            height=3cm,
            width=9cm,
            enlarge x limits=0.25,
            ylabel={Percentage of Issues},
            ylabel style={font=\small},
            symbolic x coords={Basic Type Conversion, Complex Data Structure, Missing Type Validation},
            xtick=data,
            ymin=0, ymax=60,
            nodes near coords={\pgfmathprintnumber\pgfplotspointmeta\%},
            every node near coord/.append style={font=\footnotesize},
            legend style={at={(0.5,1.05)},anchor=south,legend columns=-1},
            xticklabel style={font=\small},
            xtick pos=both,
            xtick align=inside,
            xticklabels={\shortstack{Basic Type\\Conversion}, \shortstack{Complex Data\\Structure}, \shortstack{Missing Type\\Validation}}
        ]
        \addplot[fill=black!40!white] coordinates {
            (Basic Type Conversion, 22.2) 
            (Complex Data Structure, 44.4) 
            (Missing Type Validation, 27.8) 
        };
        \end{axis}
    \end{tikzpicture}
    \vspace{-0.1in}
    \caption{Distribution of fixed patches across subcategories}
    \label{fig:sub-cat}
\end{figure}
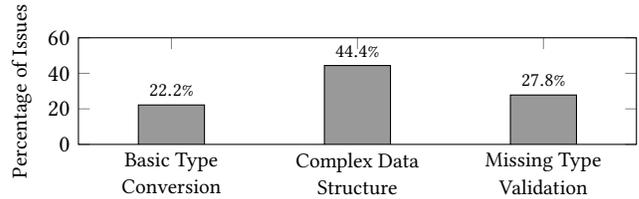

Our analysis revealed that {\tool}  successfully resolved 18 unique issues across the 29 total occurrences (with some issues appearing in multiple resolution agent's outputs). 
The distribution of these 18 unique issues across the three sub-categories is as shown in Fig.\ref{fig:sub-cat}. Complex Data Structure Manipulations represented the largest category of successfully resolved issues, accounting for 44.4\% of \tool's unique fixes. These included challenging problems like the handling of matrix symbols in sympy's common subexpression elimination (sympy\_\_sympy-22840), resolving GROUP BY clause ambiguities in django queries (django\_\_django-12589), and managing polynomial term ordering in LATEX output (sympy\_\_sympy-14317). 
{\tool} demonstrates particular strength in inferring correct type relationships within complex nested data structures and understanding how these relationships impact program behavior. Missing Type validations accounted for 27.8\% of unique successful fixes, highlighting PAGENT's ability to recognize when critical type checks were absent. Basic Type Conversions represented 22.2\% of fixed unique issues, where {\tool}  addressed fundamental type handling problems. 

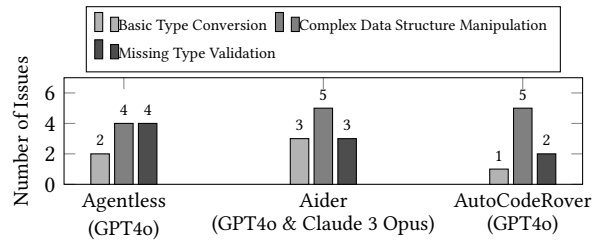
\begin{figure}[t!]
    \centering
    \begin{tikzpicture}
        \begin{axis}[
            ybar,
            bar width=7pt,
            height=3cm,
            width=\columnwidth,
            enlarge x limits=0.15,
            ylabel={Number of Issues},
            ylabel style={font=\small},
            symbolic x coords={Agentless, Aider, AutoCodeRover},
            xtick=data,
            ymin=0, ymax=7,
            nodes near coords,
            every node near coord/.append style={font=\scriptsize},
            legend style={at={(0.5,1.05)},anchor=south,legend columns=2, font=\scriptsize,
                column sep=-0.08cm,
                legend cell align=left
            },
            xticklabel style={font=\small},
            xtick align=inside,
            xticklabels={\shortstack{Agentless\\(GPT4o)}, \shortstack{Aider\\(GPT4o \& Claude 3 Opus)}, \shortstack{AutoCodeRover\\(GPT4o)}},
            legend entries={Basic Type Conversion, Complex Data Structure Manipulation, Missing Type Validation}
        ]
        \addplot[fill=black!30!white] coordinates {
            (Agentless, 2)
            (Aider, 3)
            (AutoCodeRover, 1)
        };
        \addplot[fill=black!50!white] coordinates {
            (Agentless, 4)
            (Aider, 5)
            (AutoCodeRover, 5)
        };
        \addplot[fill=black!70!white] coordinates {
            (Agentless, 4)
            (Aider, 3)
            (AutoCodeRover, 2)
        };
        \end{axis}
    \end{tikzpicture}
    \vspace{-0.1in}
    \caption{Number of fixed patches in each subcategory}
    \label{fig:subcategory-across-model}
\end{figure}

We also find that the issues that {\tool}  successfully resolved across multiple issue resolution agents predominantly fell into the complex data structure manipulation category, as shown in Fig. \ref{fig:subcategory-across-model}. This suggests that current LLM-based approaches have particular difficulty with sophisticated data structure operations where multiple types interact in complex ways. \tool's effectiveness across these sub-categories demonstrates its versatility in addressing the type-related challenges. 

{\centering
\begin{tcolorbox}
[width=1\columnwidth, boxrule=0.5pt, colback=gray!10, arc=4pt,
                  left=6pt, right=6pt, top=6pt, bottom=6pt, boxsep=0pt]
    \textbf{RQ3 Summary.}
    {\tool} resolved 18 unique issues across 29 total occurrences, with complex data structure manipulation representing the largest subcategory, 44.4\% of unique fixes, which demonstrates particular strength in handling nested data structures and inferring correct type relationships.

\end{tcolorbox}
}
\section{Discussion}
\subsection{Anatomy of Unfixed Failed Patches}

While {\tool} demonstrated some improvements across multiple type-related issue categories, it's important to analyze the cases where it failed to generate successful fixes. Through our evaluation, we observed three key categories of limitations that provide valuable insights into our approach. 

The most prevalent challenge involved complex architectural dependencies that extended beyond direct type inference. In these cases, {\tool} correctly identified the type-related issues but failed to resolve them due to an insufficient understanding of deeper architectural constraints. For example, in django\_\_django-14155, {\tool} correctly identified the need for proper handling of partial functions but couldn't generate an appropriate fix because it required understanding how the ResolverMatch class interacted with Django's URL resolution system across multiple modules. These failures highlight limitations in reasoning about complex multi-module interactions where type information alone is insufficient.

A second significant category involved domain-specific API knowledge limitations. In several computing libraries like sympy\_\_
sympy-13437, {\tool} struggled with specialized mathematical APIs that required deep domain knowledge beyond type signatures. 
In this example, {\tool} correctly identified the type inconsistency but failed to implement the appropriate domain-specific validation rules needed for correct mathematical behavior. This points to a fundamental limitation in automated approaches when domain-specific knowledge becomes combined with type handling.


Finally, {\tool} struggled with complex conditional logic around type handling, particularly when multiple type conversions needed to occur based on conditional factors. In scikit-learn\_\_scikit-learn-25638, while {\tool} identified the need for improved handling of pandas nullable types, it could not generate correct patch logic for handling the intricate branching conditions required for proper conversion across different numerical type scenarios. These cases demonstrate that even with accurate type inference, generating correct logical structures for complex type handling remains a significant challenge for automated approaches.

\subsection{Threats To Validity}
\label{sec:threats}

While we have taken significant measures to ensure the robustness of our study, several potential threats to validity warrant consideration. Our taxonomy development partially relied on LLM-based analysis, which, despite being manual verification, may introduce some categorization bias due to the inherent limitations in code semantics understanding. Regarding external validity, our focus on seven models and the SWE-bench dataset, while diverse, may not fully represent all possible issue resolution contexts or software projects. The sample size of 114 failed issues, though comprehensive for common failures across the evaluated models, could limit statistical power. 
\section{Related Work}
\label{sec:related}
\subsection{\textbf{Agent-based Issue Resolution}}
Automated Program Repair (APR) has become a pivotal area of research in software engineering~\cite{fan2023automatedrepairprogramslarge, 10172803,lecong2024semanticguidedsearchefficientprogram, jiang2023impactcodelanguagemodels, zhang2024autocoderoverautonomousprogramimprovement, xia2024agentlessdemystifyingllmbasedsoftware, moatless-tools, yang2024sweagentagentcomputerinterfacesenable, getappmap2024specification, gauthier2024aider},
aiming to automate the detection and correction of software defects to enhance system reliability and reduce manual debugging efforts. Recent advancements have introduced agent-based approaches that leverage large language models (LLMs) to autonomously plan and execute bug fixes. For instance, RepairAgent \cite{bouzenia2024repairagentautonomousllmbasedagent} utilizes an LLM to gather information about bugs, generate repair strategies, and validate fixes, demonstrating the potential of agents to handle complex repair tasks with minimal human intervention. Another approach is ChatRepair \cite{xia2024automated}, which introduces a fully automated, conversation-driven framework for program repair. Chatrepair interleaves patch generation with immediate feedback, enabling dynamic interaction between the agent and the codebase to iteratively refine fixes. Another innovative framework is FixAgent \cite{lee2024unifieddebuggingapproachllmbased}, which employs a unified debugging approach through the collaboration of multiple LLM-based agents. It addresses challenges such as implementing designs inspired by human debugging practices, including agent specialization and synergy, key variable tracking, and comprehensive program context comprehension.

Several cutting-edge issue resolution agents have emerged in this field, particularly on the SWE-bench dataset. AutoCodeROver \cite{zhang2024autocoderoverautonomousprogramimprovement} integrates GPT4o with AST representations and Spectrum-based Fault Localization (SBFL), delivering a systematic approach to patch generation. Agentless \cite{xia2024agentlessdemystifyingllmbasedsoftware} takes a fundamentally different approach by avoiding agents entirely. It uses a simple seven-phase process of localization, repair, and patch validation to fix software bugs effectively. 
Aider \cite{gauthier2024aider} employs a repository map created through static analysis of the code's AST and call graph. It uses a unique approach to code editing with specialized edit format instructions tailored to each LLM's capabilities. AppMap Naive \cite{getappmap2024specification} employs a deterministic seven-phase approach. The model is particularly notable for its ability to process full-stack applications and handle complex runtime behaviors while maintaining high standards for code quality through automated repair processes. SWE-Agent \cite{yang2024sweagentagentcomputerinterfacesenable} is one of the first end-to-end LLM agent frameworks designed specifically for repository-level software engineering. It implements a custom agent-computer interface (ACI) that allows the LLM agent to interact with the repository environment through actions like reading, editing files, and running bash commands. 
These agent-based approaches signify a transformative shift in automated program repair, leveraging the capabilities of LLMs to identify and analyze software defects. 

\subsection{\textbf{Patch Validation}}
In recent years, significant research has focused on automated approaches for software issue resolution, particularly in patch generation and validation~\cite{10.1145/3664646.3664770, 10.1145/3180155.3180233, 10.1145/3650212.3680381, 10.1145/3324884.3416590, 10440574}. Analyzing the effectiveness of patches generated by automated approaches is crucial for advancing software engineering solutions, as it provides direct evidence of an agent's problem-solving capabilities and reveals areas for improvement~\cite{zhou2024patchzerozeroshotautomaticpatch}. A fundamental challenge in this domain was highlighted by some researchers' analysis of patch plausibility and correctness across multiple automated program repair systems, including GenProg, RSRepair, and AE~\cite{10.1145/2771783.2771791}. Their study revealed that only a small fraction of generated patches were truly correct, with most patches simply deleting functionality rather than providing genuine fixes. This limitation in patch quality is further addressed by BugBuilder~\cite{bugbuilder}, which tackles the challenge of extracting clean, bug-fixing patches from version control systems. 

Recently, LLM-based approaches have been proposed to validate patches. 
Otter~\cite{ahmed2025ottergeneratingtestsissues} was introduced with an LLM-based system that generates tests from software issues before patch creation. 
More recently, FixCheck~\cite{10638611} was developed by combing random testing and large language models (LLMs) to assess patch correctness and generate fault-revealing tests, successfully identifying issues in 62\% of incorrect patches. 
\section{Conclusion}
\label{sec:conclusion}

We reported an empirical study of failed patches from issue resolution agents. By analyzing 114 failed issues across seven agents on the SWE-Bench Lite dataset, we developed a taxonomy of six distinct failure patterns. Based on these insights, we introduced {\tool}, a novel approach that enhances existing agent-based methods by addressing type-related limitations through targeted static analysis and type inference. Our evaluation demonstrates that {\tool} successfully fixed 29 previously failed patches, achieving a 9.37\% improvement rate on type-related issues. These results highlight the effectiveness of specialized processes in addressing specific limitations of general LLM-based issue resolution agents while also revealing areas where further work is needed to address more complex architectural and domain-specific challenges.

\bibliographystyle{ACM-Reference-Format}

\bibliography{paper}


\end{document}